\documentclass[graybox]{svmult}

\usepackage{mathptmx}       % selects Times Roman as basic font
\usepackage{helvet}         % selects Helvetica as sans-serif font
\usepackage{courier}        % selects Courier as typewriter font
\usepackage{type1cm}        % activate if the above 3 fonts are
\usepackage[export]{adjustbox}
\usepackage{graphicx}
\usepackage{epsfig}
\usepackage{bm}
\usepackage{amssymb}
\usepackage{float}
\usepackage{amsmath}
\usepackage{dcolumn}

\newcommand{\HCd}{\mathcal{H}}
\newcommand{\HCdtD}{\tilde{\HCd}_{\mathrm{Dyn}}}

\def\HCdtD{\tilde{\HCd}_{\mathrm{Dyn}}}

\def\HCdt0{\tilde{\HCd}_{0}}

\newcommand{\onehalf}{{\textstyle\frac{1}{2}}}

\newcommand{\pfrac}[2]{\frac{\partial{#1}}{\partial{#2}}}

\usepackage{tensor}
\usepackage{mathbbol}
\begin{document}

\title*{Gauge theory of Gravity based on the correspondence between the $1^{st}$ and the $2^{nd}$ order formalisms}
\author{D.~Benisty, E.~I.~Guendelman and J.~Struckmeier}
\institute{Physics Department, Ben-Gurion University of the Negev, Beer-Sheva 84105, Israel and Frankfurt Institute for Advanced Studies (FIAS), Ruth-Moufang-Strasse~1, 60438 Frankfurt am Main, Germany}
\maketitle
\abstract{This is a shortened version of an invited talk at the XIII International Workshop “Lie Theory
and its Applications in Physics”, June 17-23, Varna, Bulgaria. A covariant canonical gauge theory of gravity free from torsion is studied. Using a metric conjugate momentum and a connection conjugate momentum, which takes the form of the Riemann tensor, a gauge theory of gravity is formulated, with form-invariant Hamiltonian. By the metric conjugate momenta, a correspondence between the Affine-Palatini formalism and the metric formalism is established. For, when the dynamical gravitational Hamiltonian $\tilde{H}_{Dyn}$ does not depend on the metric conjugate momenta, a metric compatibility is obtained from the equation of motions, and the equations of motion correspond to the solution is the metric formalism.}

\section{Introduction}
General Relativity is one of the well tested theories in physics, with many excellent predictions. A search of a rigorous derivation of General Relativity on the basis of the action principle and the requirement that the description of any system should be form-invariant under general space time transformations has  been constructed in the framework of the Covariant Canonical Gauge theory of Gravity. 

The Covariant Canonical Gauge theory of Gravity~\cite{Struckmeier:2017vkf,Struckmeier:2017stl} is formulated within the framework of the covariant Hamiltonian formalism of classical field theories. The latter ensures by construction that the action principle is maintained in its form requiring all transformations of a given system to be canonical. The imposed requirement of invariance of the original action integral with respect to local transformations in curved space time is achieved by introducing additional degrees of freedom, the gauge fields. In the basis of the formulation there are two independent fields: the metric $g^{\alpha\beta}$, which contains the information about lengths and angles of the space time, and the connection $\gamma\indices{^{\lambda}_{\alpha\beta}}$, which contains the information how a vector transforms under parallel displacement. In this formulation, these two fields are assumed to be independent dynamical quantities in the action and referred to as the Affine-Palatini formalism (or the $1^{st}$ order formalism).

Using the structure of metric and the affine connection independently, with their conjugate momenta, yields a new formulation of gauge theories of gravity. In \cite{Struckmeier:2017vkf} the discussion was with the presence of torsion, and here we discuss about the formulation with no torsion from the beginning, that has a link between the metric affine and the metric formalism \cite{Benisty:2018fgu}.

\section{A basic formulation}
The Covariant Canonical Gauge theory of Gravity is a well defined formulation derived from the canonical transformation theory in the covariant Hamiltonian picture of classical field theories \cite{Struckmeier:2017vkf}. It identifies two independent fundamental fields, which form the basis for a description of gravity: the metric $g^{\alpha\beta}$ and the connection $\gamma\indices{^{\lambda}_{\alpha\beta}}$. In the Hamiltonian description, any fundamental field has a conjugate momentum: the metric conjugate momentum is $\tilde{k}^{\alpha\lambda\beta}$ and the connection conjugate momentum is $\tilde{q}\indices{_{\eta}^{\alpha\xi\beta}}$:
\begin{align}
S=\int_{R}\left(\tilde{k}^{\,\alpha\lambda\beta}\,g_{\alpha\lambda,\beta}-\onehalf\tilde{q}\indices{_{\eta}^{\alpha\xi\beta}}\gamma\indices{^{\eta}_{\alpha\xi,\beta}}-\tilde{\mathcal{H}}_0\right)d^{4}x
\label{eq:action-integral1}
\end{align}
where the ``tilde'' sign denotes a tensor density, which multiplies the tensor with $\sqrt{-g}$. As the conjugate momentum components of the fields are the duals of the complete set of the derivatives of the field , the formulation is referred to as ``covariant canonical''. A closed description of the coupled dynamics of fields and space-time geometry has been derived in \cite{Struckmeier:2017vkf}, where the gauge formalism yields:
\begin{equation}
S=\int_{R}\left(\tilde{k}^{\,\alpha\lambda\beta}\,g_{\alpha\lambda;\beta}-\onehalf\tilde{q}\indices{_{\eta}^{\alpha\xi\beta}}R\indices{^{\eta}_{\alpha\xi\beta}}-\HCdtD (\tilde{q},\tilde{k},g)\right)d^{4}x
\label{eq:action-integral4}
\end{equation}
As a result of the gauge procedure, all partial derivatives of tensors in Eq.~(\ref{eq:action-integral1}) reappear as covariant derivatives. The partial derivative of the (non-tensorial) connection changes into the tensor $R\indices{^{\eta}_{\alpha\xi\beta}}$, which was shown to be the Riemann-Christoffel curvature tensor:
\begin{equation}\label{eq:riemann-tensor}
R\indices{^{\eta}_{\alpha\xi\beta}}=\pfrac{\gamma\indices{^{\eta}_{\alpha\beta}}}{x^{\xi}}-
\pfrac{\gamma\indices{^{\eta}_{\alpha\xi}}}{x^{\beta}}+
\gamma\indices{^{\tau}_{\alpha\beta}}\gamma\indices{^{\eta}_{\tau\xi}}-
\gamma\indices{^{\tau}_{\alpha\xi}}\gamma\indices{^{\eta}_{\tau\beta}}.
\end{equation}
The ``dynamics'' Hamiltonian $\tilde{H}_{Dyn}$ --- which is supposed to describe the dynamics of the free (uncoupled) gravitational field --- is to be built from a combination of the metric conjugate momentum, the connection conjugate momentum, and the metric itself.
\section{A correspondence between the $1^{st}$ and the $2^{nd}$ order formalism}
In addition to the foundations of the gauge theory of gravity, it turned out that the part of the action: $\tilde{k}^{\alpha\beta\gamma}  g_{\alpha\beta;\gamma}$, which contains the metric conjugate momentum, has a strong impact as a connector between the affine-Palatini formalism (or the $1^{st}$ order formalism) and the metric formalism (or the $2^{nd}$ order formalism): 
\begin{equation}\label{t}
\mathcal{L}(g,\gamma)\,_{1^\textbf{order}} + k^{\alpha\beta\gamma}  g_{\alpha\beta;\gamma} \Leftrightarrow \mathcal{L}(g)\,_{2^\textbf{order}}
\end{equation}
In the $1^{st}$ order formalism, one assumes that there are two independent fields: the metric $g^{\mu\nu}$ and the connection $\gamma^{\mu}_{\alpha\beta}$. In contrast to that, in the $2^{nd}$ order formalism the connection is assumed to be the Levi Civita or Christoffel symbol:
\begin{equation}
\gamma\indices{^{\rho}_{\mu\nu}} = \left\{ \genfrac{}{}{0pt}{}{\rho}{\mu \nu} \right\} = \frac{1}{2} g^{\rho\lambda} (g_{\lambda\mu,\nu}+g_{\lambda\nu,\mu}-g_{\mu\nu,\lambda})
\end{equation}
and appears in the action directly in this way. In general, only for Lovelock theories, which includes Einstein Hilbert action, both formulations will yield the same equations of motion and the connection will be in both cases the Christoffel symbol. In Ref \cite{Benisty:2018fgu}, it was proved that for any general action which starts in the $1^{st}$ order formalism in addition to the term $k^{\alpha\beta\gamma}  g_{\alpha\beta;\gamma}$ the energy momentum tensor will be the same as it would be calculated in the $2^{nd}$ order formalism. The main reason for that correspondence is the metric compatibility constraint. The variation with respect to $k^{\alpha\beta\gamma}$ gives the metricity condition:
\begin{equation}\label{met}
g_{\alpha\beta;\gamma} = 0 \quad \Rightarrow \quad \gamma\indices{^{\rho}_{\mu\nu}} = \left\{ \genfrac{}{}{0pt}{}{\rho}{\mu \nu} \right\},
\end{equation}
which cause the connection to be the Christoffel symbol.  The variation with respect to the connection gives the tensors:
\begin{equation}\label{g}
\frac{\delta}{\delta \gamma^{\rho}_{\mu\nu}}  k^{\alpha\beta\gamma}  g_{\alpha\beta;\gamma} = - k^{\alpha\mu\nu}g_{\rho\alpha} - k^{\alpha\nu\mu}g_{\rho\alpha}
\end{equation}
with a symmetrization between the components $\mu$ and $\nu$. The variation with respect to the metric is:
\begin{equation}\label{eomm}
\frac{\delta}{\delta g_{\mu\nu}}  k^{\alpha\beta\gamma}  g_{\alpha\beta;\gamma} = -k\indices{^{\mu\nu\lambda}_{;\lambda}}.
\end{equation}
Because of the new contribution to the field equation $k\indices{^{\mu\nu\lambda}_{;\lambda}}$, the complete field equation will contains additional terms which make the first order field equations to be equivalent to the field equation under the second order formalism. Indeed, isolating the tensor $k^{\mu\nu\lambda}$ and inserting it back into Eq.~(\ref{eomm}) gives the relation:  
\begin{equation}\label{connection}
\frac{\partial \mathcal{L}(\kappa)}{\partial g_{\sigma\nu}}  = \frac{1}{2}\nabla_{\mu}\left(g^{\rho\sigma}\frac{\partial \mathcal{L}(\kappa)}{\partial \gamma\indices{^{\rho}_{\mu\nu}}}+g^{\rho\nu}\frac{\partial \mathcal{L}(\kappa)}{\partial \gamma\indices{^{\rho}_{\mu\sigma}}}-g^{\rho\mu}\frac{\partial \mathcal{L}(\kappa)}{\partial \gamma\indices{^{\rho}_{\nu\sigma}}}\right)
\end{equation}
where $\mathcal{L}(\kappa) = k^{\alpha\beta\gamma}  g_{\alpha\beta;\gamma}$.
The terms in the right hand side represents the additional terms that appear in the second order formalism. One option for obtain the contributions into the field equation is to solve $k^{\alpha\beta\gamma}$. The direct way is by using this equation, that gives the new contributions for the second order formalism into the field equation, from the variation with respect to the connection $\gamma^{\rho}_{\mu\nu}$. An application for this correspondence is from the Covariant Canonical Gauge theory of gravity action (\ref{eq:action-integral4}).
\section{Path to Gauge Theories}
From the correspondence between the $1^{st}$ and the $2^{nd}$ order formalisms theorem, we obtain a basic link between the dependence of the $\mathcal{H}_{\textbf{Dyn}}$ with the metric conjugate momentum $\tilde{k}^{\alpha\beta\gamma}$ and the structure of the metric energy momentum tensor. In the first case, $\mathcal{H}_{\textbf{Dyn}}$ does not depend on the metric conjugate momentum $\tilde{k}^{\alpha\beta\gamma}$.
A variation with respect to the metric conjugate momentum $\tilde{k}^{\alpha\beta\gamma}$ gives the metric compatibility condition. According to the theorem (\ref{t}) the gravitational energy momentum tensor is the same as the gravitational energy momentum tensor in the second order formalism. In the second case $\mathcal{H}_{\textbf{Dyn}}$ does depend on the metric conjugate momentum $\tilde{k}^{\alpha\beta\gamma}$. A variation with respect to the metric conjugate momentum $\tilde{k}^{\alpha\beta\gamma}$ breaks the metric compatibility condition. This basic framework is not a special feature only for the Covariant Canonical Gauge Theory of Gravity, but leads to a fundamental correlation for many options of $\mathcal{H}_{Dyn}$.

In analogy to the definition of the metric energy-momentum tensor density of the given system Hamiltonian, the metric energy-momentum tensor density is being  define as the variation of the $\tilde{\mathcal{L}}_m$ with respect to the metric:
\begin{equation}
T^{\mu\nu} = -\frac{2}{\sqrt{-g}}\frac{\partial\tilde{\mathcal{L}}_m}{\partial g_{\mu\nu}}
\end{equation}
Therefore the complete action takes the form:
\begin{equation}\label{sec:intro}
\tilde{\mathcal{L}} = \tilde{k}^{\alpha\beta\gamma} g_{\alpha\beta;\gamma} - \frac{1}{2} \tilde{q}\indices{_{\lambda}^{\alpha\beta\gamma}}R\indices{^{\lambda}_{\alpha\beta\gamma}}-\tilde{\mathcal{H}}_{\textbf{dyn}} (\tilde{q},\tilde{k},g) + \mathcal{\tilde{L}}_m 
\end{equation}
The variation with respect to the metric conjugate momenta:
\begin{equation}
g_{\alpha\beta;\gamma} = \frac{\partial  \tilde{H}_{\textbf{Dyn}}}{\partial \tilde{k}^{\alpha\beta\gamma}}  
\end{equation}
which presents the existence of non-metricity if $\mathcal{H}_{Dyn}$ depends on $k^{\alpha\beta\gamma}$. The second variation is the variation with respect to the connection:
\begin{equation}\label{varCon}
-\left(\tilde{k}^{\alpha\mu\nu}+\tilde{k}^{\alpha\nu\mu}\right) g_{\alpha\rho} = \frac{1}{2}\nabla_{\beta} \left( \tilde{q}\indices{_{\rho}^{\mu\beta\nu}}+ \tilde{q}\indices{_{\rho}^{\nu\beta\mu}} \right),
\end{equation}
which contracts the relation between the momenta of the metric and the connection.
The third variation is with respect to the connection conjugate momentum $\tilde{q}\indices{_{\sigma}^{\mu\nu\rho}}$, which gives:
\begin{equation}
\frac{\partial \tilde{H}_{\textbf{Dyn}}}{\partial \tilde{q}\indices{_{\sigma}^{\mu\nu\rho}}} = -\frac{1}{2} R\indices{^{\sigma}_{\mu\nu\rho}}
\end{equation}
If $\tilde{H}_{\textbf{Dyn}}$ is not depend on $\tilde{q}$, the Riemann tensor will be zero. Therefore the contribution for the stress energy tensor comes from the Dynamical Hamiltonian and from the metric conjugate momenta:
\begin{equation}
\begin{split}
T^{\mu\nu} =g^{\mu\nu}(k^{\alpha\beta\gamma} g_{\alpha\beta;\gamma} - \frac{1}{2} q_{\lambda}^{\alpha\beta\gamma}R^{\lambda}_{\alpha\beta\gamma})-2k^{\mu\nu\gamma}_{;\gamma} + \frac{2}{\sqrt{-g}}\frac{\partial\tilde{\mathcal{H}}_\textbf{Dyn}}{\partial g_{\mu\nu}}
\end{split}
\end{equation}
From the variation with respect the connection (\ref{varCon}), the value of the momentum $\tilde{k}^{\alpha\beta\gamma}$.
As an example, we consider a Dynamical Hamiltonian which has no dependence with the metric conjugate momentum. 
\section{Sample $\tilde{\mathcal{H}}_{\textbf{dyn}}$ without breaking metricity}
Our starting point is a dynamical Hamiltonian with the connection conjugate momentum up to the second order, without a dependence on the metric conjugate momentum:
\begin{equation}
\tilde{\mathcal{H}}_{\textbf{dyn}} = \frac{1}{4g_1} \tilde{q}\indices{_{\eta}^{\alpha\epsilon\beta}}q\indices{_{\alpha}^{\eta\tau\lambda}} g_{\epsilon\tau} g_{\beta\lambda} - g_2 \, \tilde{q}\indices{_{\eta}^{\alpha\tau\beta}} g_{\alpha\beta}\delta^{\eta}_{\tau} + g_3 \sqrt{-g}
\end{equation}
This Hamiltonian was investigated in \cite{Vasak:2018gqn} under the original formalism for non-zero torsion (which is finally set to zero), and led to resolving the cosmological constant problem. In our case, assuming that there is no torsion, the formalism demands that the energy momentum tensor is covariantly conserved, as is supposed to be in the second order formalism. 

The variation with respect to the metric conjugate momenta $\tilde{k}^{\alpha\beta\gamma}$ give the the metricity condition. The variation with respect to the connection conjugate momenta $\tilde{q}\indices{_{\sigma}^{\mu\nu\rho}}$ gives
\begin{equation}\label{qdyn}
q_{\eta\alpha\epsilon\beta} = g_1 \left( R_{\eta\alpha\epsilon\beta} - \hat{R}_{\eta\alpha\epsilon\beta} \right)
\end{equation}
where: 
\begin{equation}
\hat{R}_{\eta\alpha\epsilon\beta} = g_2 \left(g_{\eta\epsilon}g_{\alpha\beta} -g_{\eta\beta}g_{\epsilon\alpha} \right)
\end{equation}
refers to the ground state geometry of space-time which is the de Sitter (dS) or the anti-de Sitter (AdS) space-time for the positive or the negative sign of $g_2$, respectively.
The last variation is with respect to the metric. 
In order to isolate the tensor $k^{\mu\nu\gamma}$ one can use the following process: First, we multiply by the metric $g^{\rho\sigma}$ and sum over the index $\sigma$:
\begin{equation}\label{varCon1}
-\tilde{k}^{\sigma\mu\nu}-\tilde{k}^{\sigma\nu\mu}= \frac{1}{2}\nabla_{\alpha} \left( \tilde{q}^{\,\,\sigma\mu\alpha\nu}+ \tilde{q}^{\,\,\sigma\nu\alpha\mu} \right)
\end{equation}
Switching the indices $\sigma \leftrightarrow \nu$ and the indices $\mu\leftrightarrow \nu$ gives a new combination of the $k^{\alpha\beta\gamma}$ tensor. By summing the Eqs. the isolated value gives:
\begin{equation}\label{varCovf}
-\tilde{k}^{\sigma\nu\mu} = \frac{1}{2} \nabla_\alpha \left(\tilde{q}^{\sigma\mu\alpha\nu}+\tilde{q}^{\nu\mu\alpha\sigma}\right)
\end{equation}
Therefore, the contribution for the stress energy momentum comes from the covariant derivative of (\ref{varCovf}):
\begin{align}
T^{\mu\nu} &= - \frac{1}{2} g^{\mu\nu}q\indices{_{\lambda}^{\alpha\beta\gamma}}R\indices{^{\lambda}_{\alpha\beta\gamma}} + \nabla_{\gamma}\nabla_{\alpha}\left(q^{\mu\gamma\nu\alpha} + q^{\nu\gamma\mu\alpha}\right) + \frac{2}{\sqrt{-g}}\frac{\partial\tilde{\mathcal{H}}_\textbf{Dyn}}{\partial g_{\mu\nu}}
\end{align}
Plugging in the value of the tensor $q_{\alpha\beta\gamma\delta}$ from Eq.~(\ref{qdyn}) gives the result:
\begin{equation}
\begin{split}
T^{\mu\nu} = \frac{1}{8 \pi G}G^{\mu\nu} +g^{\mu\nu} \Lambda \\ + g_1 \left(R^{\mu\alpha\beta\gamma}R^{\nu}_{\alpha\beta\gamma} - \frac{1}{4} g^{\mu\nu} R^{\alpha\beta\gamma\delta}R_{\alpha\beta\gamma\delta} + (\nabla_{\alpha}\nabla_{\beta} + \nabla_{\beta}\nabla_{\alpha}) R^{\mu\alpha\nu\beta} \right)
\end{split}
\end{equation}
where $G^{\mu\nu}$ is the Einstein tensor. The coupling constants relate to the physical quantities with the relations: 
\begin{equation}
g_1 g_2 = \frac{1}{16 \pi G}, \quad 6g_1 g_2^2 + g_3 = \frac{\Lambda}{8 \pi G}
\end{equation}
This stress energy momentum tensor is exactly the same metric energy momentum tensor if our starting point was the effective Lagrangian:
\begin{equation}\label{action}
\mathcal{L} = g_1 R^{\alpha\beta\gamma\delta} R_{\alpha\beta\gamma\delta} - \frac{1}{16\pi G} (R - 2\Lambda)  
\end{equation}
and the stress energy momentum tensor is the same stress energy momentum tensor for this Lagrangian in the second order formalism. One from the big benefits of this formulation is common in many gauge theories of gravity \cite{Lagraa:2016knm}, where the starting point is with additional variables with no higher derivatives in the action, and the equations of motion are equivalent to actions with higher derivatives of metric. In this specific case, the starting point is with the quartic momentum $q$ and at the end is equivalent to an action with quadratic Riemann term.  
\section{Discussion}
In this paper we investigated the formulation of the covariant canonical gauge theory of gravity free from torsion. Diffeomorphisms appear as canonical transformations. A tensor field which plays the role of the canonical conjugate of the metric is introduced. It enforces the metricity condition provided that the ``Dynamics'' Hamiltonian does not depend on this field. The resulting theory has a direct correspondence with our recent work concerning the correspondence between the first order formalism and the second order formalism through the introduction of a Lagrange multiplier field which in this case corresponds with the field that is used to provide the metric with a canonically conjugate momentum. The procedure is exemplified by using a ``Dynamics'' Hamiltonian which consists of a quadratic term of the connection conjugate momentum. The effective stress energy momentum tensor that emerged from the canonical equations of motion were equivalent to Einstein Hilbert tensor in addition to quadratic Riemann term. \cite{Benisty:2018ufz} derives the complete combination for the quadratic theories of gravity with $R^2$ and $R_{\mu\nu}R^{\mu\nu}$.   \cite{Benisty:2019jqz} derives inflation from fermios based on the Covariant Canonical Gauge theories of Gravity approach to spinors \cite{Struckmeier:2018psp}.

\begin{acknowledgement}
We gratefully acknowledge support of our collaboration through the Exchange Agreement between Ben-Gurion University, Beer-Sheva, Israel and Bulgarian Academy of Sciences, Sofia, Bulgaria. D.B. partially supported by COST Actions CA15117, CA16104 and the action CA18108.
\end{acknowledgement}

\end{document}